\documentclass{PoS}
\usepackage{epsfig}

\usepackage{amsmath}

\usepackage{amssymb}
\usepackage{amsbsy}
\usepackage{amsfonts}
\usepackage{graphics}
\usepackage{latexsym}
\usepackage{mathrsfs}
\usepackage{dcolumn}
\usepackage{wasysym}

\title{
\vspace{-1cm}
\begin{minipage}{\textwidth}
\begin{flushright}
\normalsize DESY-11-026  \\ Edinburgh 2011/10 \\ ITEP-LAT/2011-02 \\ 
\end{flushright}
\end{minipage}\\[15pt]
Finite temperature phase transition with two flavors of 
improved Wilson fermions}

\ShortTitle{Phase transition}

\author{\speaker{V.G.~Bornyakov} \\
        Institute for high energy physics, 142281 Protvino, Russia\\
        E-mail: \email{vitaly.bornyakov@ihep.ru}}

\author{R. Horsley \\
       School of Physics and Astronomy, University of Edinburgh, Edinburgh
EH9 3JZ, UK\\}
\author{ Y. Nakamura\\
        Institut f\"ur Theoretische Physik, Universit\"at Regensburg,
93040 Regensburg, Germany\\}
\author{ M. Polikarpov\\
        Institute of Theoretical and Experimental Physics ITEP, 117259 Moscow, Russia\\}
\author{P. Rakow  \\
        Theoretical Physics Division, Department of Mathematical Sciences,
University of Liverpool, Liverpool L69 3BX, UK\\}
\author{ G. Schierholz\\
        Deutsches Elektronen-Synchrotron DESY, 22603 Hamburg, Germany \\}
\abstract{The critical temperature is computed 
for $N_f = 2$ dynamical flavors of nonperturbatively improved Wilson
fermions. The new simulations are performed on lattices $40^3 x 14$ with lattice
spacing and pion mass about 0.08 fm and 200 MeV, respectively.
We find the deconfinement and chiral phase transitions to coincide within numerical
precision.  Our results are in broad agreement with a second order phase
transition in the chiral limit. The critical temperature at the physical quark mass
is found to be  $T_c = 172(3)(6)\, \mbox{MeV}$.
}

\FullConference{The XXVIII International Symposium on Lattice Field Theory, Lattice2010\\
		June 14-19, 2010\\
		Villasimius, Italy}

\begin{document}
\section{Introduction}

The nature of the finite temperature phase transition, and the value of the transition 
temperature are basic questions in finite temperature QCD which are still missing 
final answeres. For QCD with 2+1 flavors one group~\cite{W} finds the deconfining
transition, and the chiral transition  temperatures are separated by 20-30 MeV while 
another group~\cite{BB2} claims both temperatures to coincide. Moreover, the 
Brookhaven/Bielefeld collaboration~\cite{BB2} 
gets for transition temperature $T_c = 196(3) \, \mbox{MeV}$,
which is much higher than the transition temperatures found by the Wuppertal
group~\cite{W2} for the deconfining and chiral transitions - 
$T_c = 170(7)\, \mbox{MeV}$,   and $T_c = 146(5)\, \mbox{MeV}$, respectively.
Both groups use rooted staggered fermions, but with
different levels of improvement. It has been
argued~\cite{W} that the discrepancy is largely due to the rather
coarse lattices used by the Brookhaven/Bielefeld collaboration. 
Recently, the Brookhaven/Bielefeld collaboration
has extended their calculations to lattices of temporal extent $N_t =
8$~\cite{BBB} and found that with decreasing lattice spacing $T_c$ was 
shifted by $5 - 7 \, \mbox{MeV}$ towards smaller values.

The connection between deconfining and chiral transition has been 
subject of several phenomenological considerations. Naively, one would
expect the temperature of the deconfinement transition to lie below that of
the chiral transition, if different at all. This turns out to
be the case, for example, in the Polyakov-loop extended Nambu--Jona-Lasinio
model~\cite{Weise}. More likely is that both transitions
occur at the same temperature, as Polyakov loop and chiral condensate mix
at finite dynamical quark masses. The consequence would be a simultaneous
enhancement of both the chiral and Polyakov-loop susceptibilities along the
transition line~\cite{mix,mix2,mix3,mix4}.

To clarify the issue, independent investigations of the nature of
the finite temperature phase transition preferably with different type 
of the lattice fermionic action are needed (for recent related works see 
\cite{Cheng:2009be} 
(domain wall), \cite{Burger:2010ag} (twisted mass) and 
\cite{Brandt:2010uw} (improved Wilson fermions) ). In this work we
present results of simulations with $N_f=2$ dynamical flavors of
nonperturbatively $O(a)$ improved Wilson fermions and plaquette gauge
action on lattices of temporal extent $N_t=14, 12, 10$ and $8$. 
Our results were partially reported in~\cite{DIK,Bornyakov:2009qh}.

\section{Definitions and simulation parameters}

The fermionic action for each of the two flavors reads
\begin{equation}
\begin{split}
 S_F =  a^4 \sum_x \,\Big\{ & \frac{1}{2 a} \,\sum_\mu \,\bar{\psi}(x) \;
 U_\mu(x)\, \left[\gamma_\mu - 1\right]\, \psi(x +a \hat{\mu}) \\
 -\; &\frac{1}{2 a}\, \sum_\mu \,\bar{\psi}(x) \;
 U^\dagger_\mu(x -a \hat{\mu})\, \left[\gamma_\mu + 1\right]\, \psi(x
 -a \hat{\mu})\\
 -\; &c_{SW} \,\frac{i}{2a} \, \sum_{\mu\nu} \,
 \bar{\psi}(x)\,\sigma_{\mu\nu}\, P_{\mu\nu}(x)\, \psi(x)
  + (m + m_c) \, \bar{\psi}(x)\, \psi(x) \Big\}
 \label{SW_action}
\end{split}
\end{equation}
 where $P_{\mu\nu}$ is the clover-leaf form of the  lattice field strength tensor and
 \begin{equation}
 a m_c = \frac{1}{2 \kappa_c} \,, \quad a m = \frac{1}{2 \kappa} -
 \frac{1}{2 \kappa_c} \,
 \end{equation}
$\kappa_c$ being the critical value of the hopping parameter. 
\begin{table}[t]
\begin{center}
\vspace*{0.25cm}
\begin{tabular}{|c|c|c|c|c|}\hline
$\beta$ & $c_{SW}$ & $V=N_s^3\, N_t$ & $\kappa_c$ & $r_0/a$ \\
\hline
5.20  & $2.0171$ & $16^3\, 8$\phantom{0} & $0.136050(17)$ & $5.454(58)$ \\
5.20  & $2.0171$ & $24^3\, 10$ & $0.136050(17)$ & $5.454(58)$ \\
5.25  & $1.9603$ & $16^3\, 8$\phantom{0} & $0.136273(7)$\phantom{0} & $5.880(26)$ \\
5.25  & $1.9603$ & $24^3\, 8$\phantom{0} & $0.136273(7)$\phantom{0} & $5.880(26)$ \\
5.25  & $1.9603$ & $32^3\, 12$ & $0.136273(7)$\phantom{0} & $5.880(26)$ \\
5.25  & $1.9603$ & $40^3\, 14$ & $0.136273(7)$\phantom{0} & $5.880(26)$ \\
5.29  & $1.9192$ & $24^3\, 12$ & $0.136440(4)$\phantom{0}& $6.201(25)$ \\\hline
\end{tabular}
\end{center}
\caption{Parameters of the simulation.}
\label{param}
\end{table}

The couplings, lattice volumes and lattice
spacings covered by our simulations are listed in Table~\ref{param}.
The scale parameters $r_0/a$ have been taken from the zero temperature
runs of the QCDSF collaboration at the corresponding couplings. They
refer to the chiral limit $\kappa = \kappa_c$. We also list the
critical hopping parameters $\kappa_c$, which we adopted from QCDSF as
well. (For recent relevant work see~\cite{QCDSF2}.) The gauge field
configurations were generated on the BlueGene/L at KEK, the RSCC
cluster at RIKEN, the MVS-100k at the Joint Computer Center (Moscow),
on the SKIF-Chebyshev at Moscow State University, as well as on the
Altix at HLRN.

Two-flavor QCD is expected to undergo a second order transition at finite
temperature in the chiral limit and at very small quark masses. In the chiral
limit the order parameter is the chiral condensate
\begin{equation}
\sigma=\frac{a^3}{V}\, \sum_{x} \bar{\psi}(x) \psi(x) \,.
\end{equation}
For heavy quark masses close to the quenched limit, the theory is known to undergo 
a first order phase transition. In that limit the order parameter is the Polyakov loop
\begin{equation}
L=\frac{1}{N_s^3}\,\sum_{\vec{x}}\, {\rm Re}\, L(\vec{x})\, , \,
L(\vec{x})=\frac{1}{3} \,
{\rm Tr}\, \prod^{N_t}_{x_4=1}U_4(x) \,.
\end{equation}

The temperature of the chiral transition is, for general $m$, identified with
the peak position of the chiral susceptibility
\begin{equation}
\chi_\sigma \equiv \langle\sigma^2\rangle_c = \langle \sigma^2\rangle - \langle
\sigma\rangle^2 \,,
\end{equation}
while the peak of the Polyakov-loop susceptibility
\begin{equation}
\chi_L \equiv  N_s^3\, \langle L^2\rangle_c \, , \, \langle
L^2\rangle_c = \left( \langle
  L^2\rangle - \langle L\rangle^2\right)
\end{equation}
defines the temperature of the deconfining transition.

It is expected that the two-flavor theory is in the same universality class 
as the three-dimensional O(4) Heisenberg model~\cite{PW}, with
the external magnetic field and the magnetization being identified
with the bare quark mass $\hat{m} \equiv am$ and the chiral condensate
$\hat{\sigma} \equiv \langle \sigma \rangle$, respectively. 
The critical exponents of this model were found to be~\cite{cex}
\begin{equation}
\frac{1}{\beta\delta} = 0.537(7) \,,\,\,\, \frac{1}{\delta} = 0.206(1) \,.
\end{equation}
Let $T_c(m)$ denote the pseudocritical temperature at finite $m$,
which we define to be the temperature corresponding to the position of the 
peak of the chiral susceptibility
\begin{equation}
\chi_\sigma = \frac{\partial\,\hat{\sigma}}{\partial\,\hat{m}} \,.
\end{equation}
From the scaling relation connecting  the chiral condensate, the
dynamical quark mass and the temperature in the vicinity of the phase transition
 we then derive
\begin{equation}
T_c(m)-T_c(m=0) \, \propto \, \hat{m}^{\frac{1}{\beta\delta}} \,.
\end{equation}
Assuming
\begin{equation}
m_\pi^2 \propto m \,,
\end{equation}
we thus expect to find
\begin{equation}
T_c(m)-T_c(m=0) \, \propto \, m_\pi^{1.07(1)}
\label{o4}
\end{equation}
for a second order transition at $m=0$. A first order transition, on the other
hand, would give
\begin{equation}
T_c(m)-T_c(m =0) \, \propto \, m_\pi^2 \,.
\label{1st}
\end{equation}

The chiral condensate is related to the average plaquette $P$ by means of a
Maxwell relation~\cite{Gockeler}.  The chiral condensate and the plaquette can be 
found from the partial derivatives of the partition function $Z$:
\begin{eqnarray}
\frac{1}{V} \left.\frac{\partial}{\partial \, \beta} \ln Z
\right|_{\hat{m}} &=& -6\, P  + 2\,\frac{\partial \, \hat{m}_c}{\partial \,
  \beta}\, \hat{\sigma} - 2\, \frac{\partial \, c_{SW}}{\partial \,
  \beta}\, \hat{\delta}\,,\label{plaquette}\\ 
\frac{1}{V} \left.\frac{\partial}{\partial \, \hat{m}} \ln Z
\right|_\beta &=& 2\, \hat{\sigma} \,,
\end{eqnarray}
where last term in (\ref{plaquette}) comes from the clover term and is neglected below
because it is suppressed by two orders of the lattice spacing with respect
to the chiral condensate~\cite{Kremer,Doi,DiGiacomo}.

The second derivative $\partial^2 \ln  Z/\partial \beta\, \partial \hat{m}$ can
be expressed in two different orders, which leads to relation:
\begin{equation}
\left.\frac{\partial \,P}{\partial \, \hat{m}}\right|_\beta -
\frac{1}{3} \, \frac{\partial \, \hat{m}_c}{\partial \, \beta}\,
\left.\frac{\partial \, \hat{\sigma}}{\partial\,
  \hat{m}}\right|_{\beta} = \frac{1}{3}
\left.\frac{\partial \,\hat{m}}{\partial \, \beta}\right|_{\hat{\sigma}} \,
\left.\frac{\partial \,\hat{\sigma}}{\partial \,\hat{m}}\right|_\beta \,.
\label{plaquette2}
\end{equation}
called the Maxwell relation. It holds for any lattice size and for all values of $\beta$ and $m$. 
Then chiral condensate susceptibility can be expressed as
\begin{equation}
\chi_\sigma = \frac{1}{\mu} \frac{\partial P}{\partial \hat{m}}\,, 
\end{equation}
where
\begin{equation}
\mu^{-1} = 3 \left(\left.\frac{\partial \,\hat{m}_c}{\partial \,
      \beta} + \frac{\partial \,\hat{m}}{\partial \,
      \beta}\right|_{\hat{\sigma}}\right)^{-1} \,
\nonumber
\end{equation}
is a finite number.

We also computed the correlator $\langle L\sigma\rangle_c$ which can be obtained
from the derivative of the average Polyakov loop with respect to mass:
\begin{equation}
\langle L\sigma\rangle_c = \left.\frac{\partial \langle L\rangle}{\partial
    \hat{m}}\right|_\beta \,.
\label{Ls}
\end{equation}

\section{Transition temperature}

\begin{figure*}[t]
\centering
\includegraphics[width=0.4\textwidth]{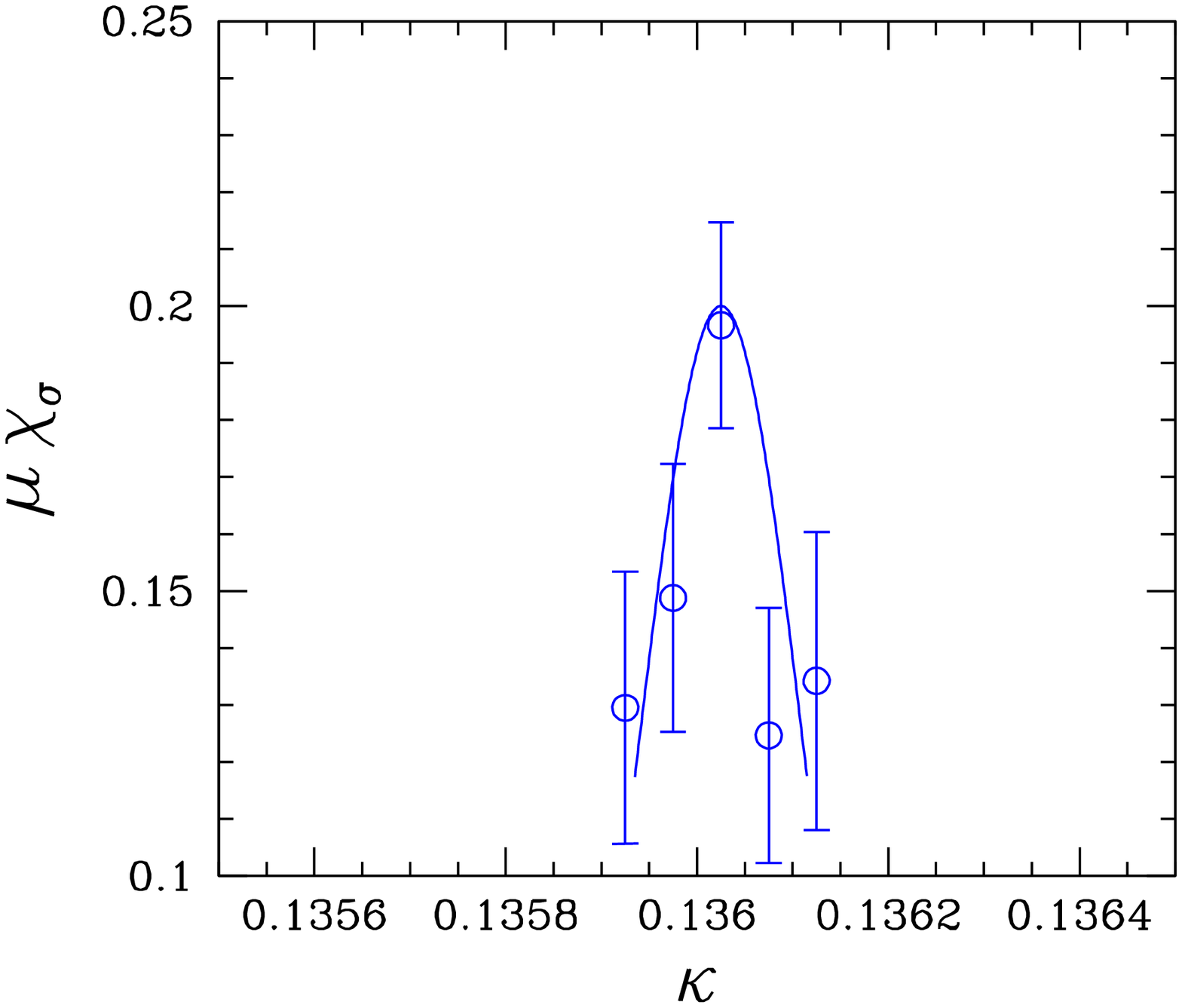}
\includegraphics[width=0.4\textwidth]{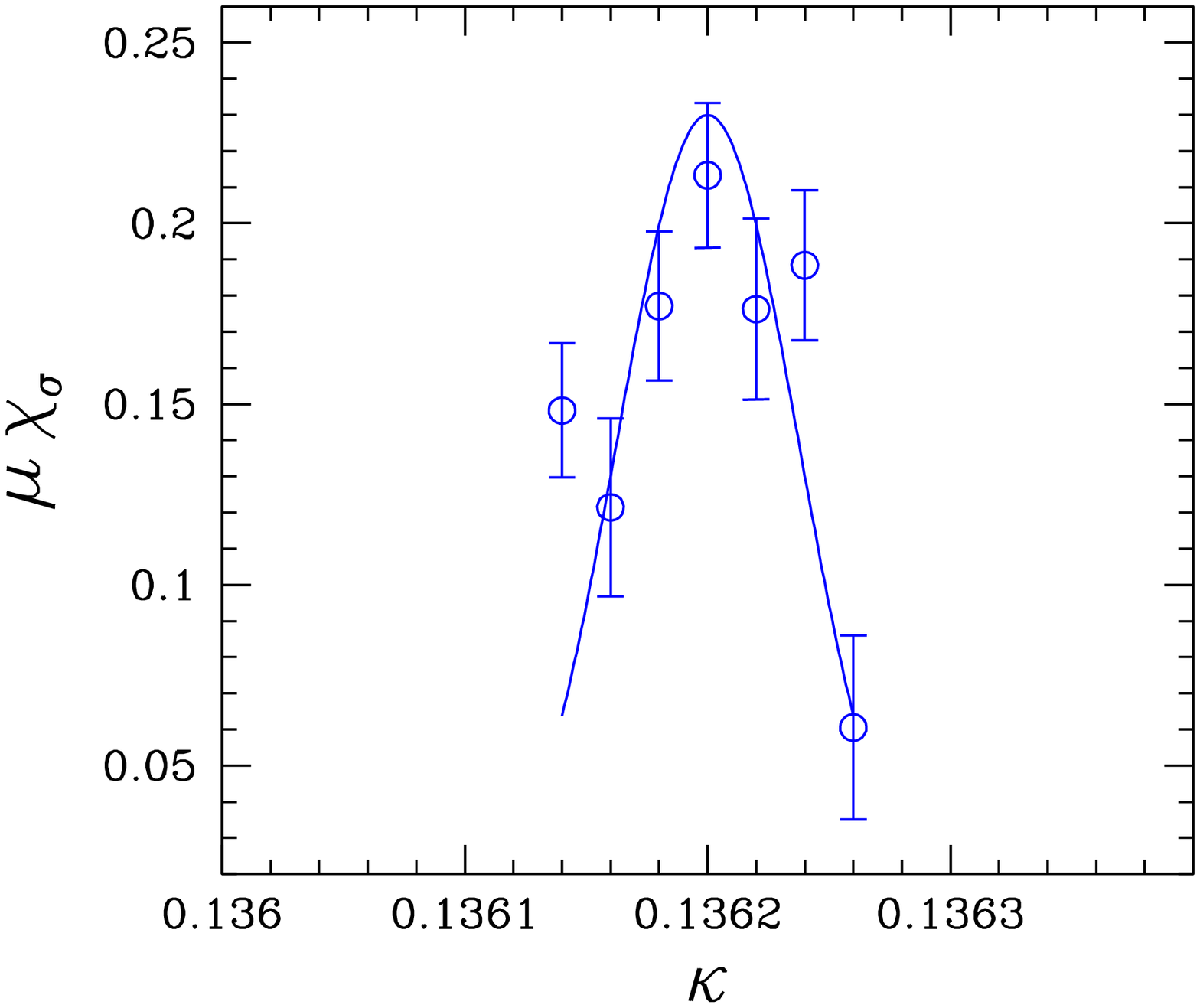}
\caption{The chiral susceptibility on the $32^3\, 12$ (left) and  $40^3\, 14$ lattice (right) at 
$\beta=5.25$ together with a Gaussian fit.}
\label{ch_susc}
\end{figure*}

In Figure~\ref{ch_susc} we show the chiral susceptibility for lattices  with $N_t=12$ and $14$ 
corresponding to our lowest quark masses. 

\begin{figure*}[t]
\centering
\includegraphics[width=0.6\textwidth]{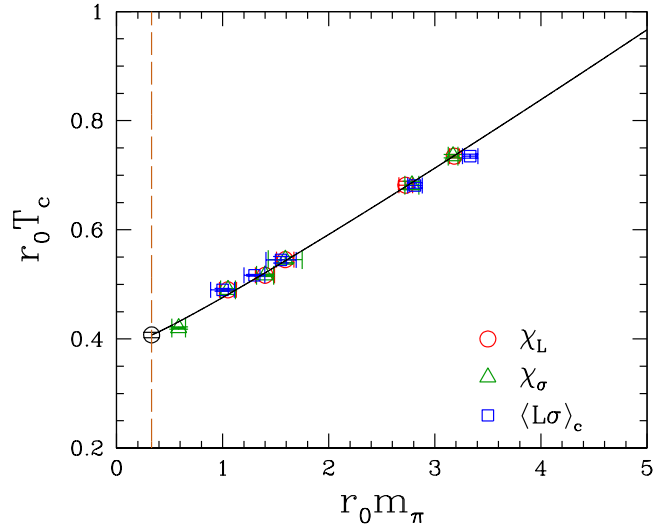}
\caption{The pseudocritical temperature $T_c(m)$ as a function of pion
  mass, together with a fit to the power $m_\pi^{1.07}$,
  according to the three-dimensional $O(4)$ model.}
\label{tcm}
\end{figure*}

In Table~\ref{kt} we show the pseudocritical
temperature $T_c(m)$ and the corresponding pseudocritical pion masses,
$m_\pi^{T_c}$, obtained from the peak of the Polyakov-loop susceptibility, 
the chiral susceptibility  and the correlator (\ref{Ls}) of $L$ and $\sigma$,
 respectively. In Fig.~\ref{tcm}  the results together with an extrapolation to 
the chiral limit are presented. We find that on all lattices the individual pion 
masses $\displaystyle m_\pi^{T_c}$ coincide with each other within the error
bars.
\begin{table}[t]
\begin{center}
\vspace*{0.25cm}
\begin{tabular}{|c|c|c|c|c|c|}\hline
 & & & \multicolumn{3}{c|}{$r_0\, m_\pi^{T_c}$} \\
\raisebox{10pt}{$\beta$} & \raisebox{10pt}{$V$} & \raisebox{10pt}{$r_0\,T_c(m)$} & $\chi_L$ & $\chi_\sigma$ & $\langle
  L\sigma\rangle_c$ \\
\hline
5.20  & $16^3\, 8$\phantom{0} & 0.682(7) & 2.73(6)\phantom{0} & 2.78(6)\phantom{0} & 2.81(7)\phantom{0} \\
5.20  & $24^3\, 10$ & 0.545(6) & 1.59(8)\phantom{0} & 1.59(16) & 1.55(14) \\
5.25  & $24^3\, 8$\phantom{0} & 0.735(3) & 3.18(4)\phantom{0} & 3.17(4)\phantom{0} & 3.33(7)\phantom{0} \\
5.25  & $32^3\, 12$ & 0.490(2) & 1.00(11)& 1.05(8)\phantom{0} & 1.05(7)\phantom{0} \\
5.25  & $40^3\, 14$ & 0.420(2) &         & 0.59(6)\phantom{0} &  \\
5.29  & $24^3\, 12$ & 0.517(2) & 1.49(8)\phantom{0} & 1.40(9)\phantom{0} & 1.3(1)\phantom{00}\\\hline
\end{tabular}
\end{center}
\caption{The pseudocritical temperatures and corresponding pion
  masses obtained from the peak of $\chi_L$, $\chi_\sigma$ and $\langle
  L\sigma\rangle_c$ on our various lattices.}
\label{kt}
\end{table}

One can see from the Figure~\ref{tcm} that the temperature
$T_c(m)$ shows an almost linear behavior in the pion mass, in
accord with the prediction (\ref{o4}) of the $O(4)$ model. We thus may
fit the data by the {\it ansatz}
\begin{equation}
T_c(m) = C + D\, (r_0\,m_\pi)^{1.07} \,.
\end{equation}
The result is shown by the solid curve. Setting the scale by the
nucleon mass, the QCDSF collaboration finds $r_0=0.467(15)\,
\mbox{fm}$. Using this value, we obtain at the physical pion mass
\begin{equation}
r_0\,T_c = 0.408(5) \; , \quad T_c=172(3)(6) \, \mbox{MeV} \,,
\end{equation}
where the first error on $T_c$ is statistical, and the second error reflects the
uncertainty in setting the scale. This result only slightly differes from our result \cite{Bornyakov:2009qh} 
obtained without $40^3\, 14$ lattice. It is in good agreement with the deconfining transition 
temperature found by the Wuppertal group, but lies significantly below the result of the 
Brookhaven/Bielefeld collaboration.

\section{Conclusions}

We have simulated QCD at finite temperature with two dynamical flavors
of nonperturbatively improved Wilson fermions on lattices as large as
$N_t=14$ and lattice spacings as low as $0.075\, \mbox{fm}$.
The transition temperature has been computed from the Polyakov-loop
susceptibility, the chiral susceptibility as well as the correlator of
Polyakov loop and chiral condensate. All three temperatures are found
to coincide with each other within the error bars. The
critical behavior appears to be in accord with the predictions of the $O(4)$
Heisenberg model, at least as far as the quark mass dependence of
$T_c$ is concerned.

Let us note that the Maxwell relation used to compute the chiral condensate 
susceptibility has proven to be a powerful tool in unveiling the
phase structure of clover fermions. 
 
\section*{Acknowledgment}
We like to thank the computer centers at KEK (under the Large Scale
Simulation Program No. 07-14-B), JSSC RAS (Moscow), 
Supercomputing Center MSU (Moscow), RIKEN and HLRN (Berlin and Hannover) 
for their generous allocation of computer time and technical support.

\end{document}